# Spatial beam self-cleaning in tapered Yb-doped GRIN multimode fiber with decelerating nonlinearity

A. Niang[1,*], D. Modotto[1], A. Tonello[2], F. Mangini[1], U. Minoni[1], M. Zitelli[3], M. Fabert[2], M.A. Jima[1,2], O.N. Egorova[4], A.E. Levchenko[5], S.L. Semjonov[5], D.S. Lipatov[6], S. Babin[7,8], V. Couderc[2], and S. Wabnitz[3,7]

[1]*Dipartimento di Ingegneria dell'Informazione, Università di Brescia, via Branze 38, 25123, Brescia, Italy*
[2]*Universitéde Limoges, XLIM, UMR CNRS 7252, 123 Avenue A. Thomas, 87060 Limoges, France*
[3]*Dipartimento di Ingegneria dell'Informazione, Elettronica e Telecomunicazioni, Sapienza University of Rome, Via Eudossiana 18, 00184 Rome, Italy*
[4]*Prokhorov General Physics Institute of the Russian Academy of Sciences, 38 Vavilov Street, 119991 Moscow, Russia*
[5]*Fiber Optics Research Center of the Russian Academy of Sciences, 38 Vavilov Street, 119333 Moscow, Russia*
[6]*Devyatykh Institute of Chemistry of High-Purity Substances of the Russian Academy of Sciences, 49 Tropinin Street, 603950 Nizhny Novgorod, Russia*
[7]*Novosibirsk State University, Pirogova 1, Novosibirsk 630090, Russia*
[8]*Institute of Automation and Electrometry of the Russian Academy of Sciences, Koptyuga 1, Novosibirsk 630090, Russia*
*alioune.niang@unibs.it

**Abstract**: We experimentally demonstrate spatial beam self-cleaning in an Yb-doped graded-index multimode fiber taper, both in passive and active configurations. The input laser beam at 1064 nm was injected for propagation from the small to the large core side of the taper, with laser diode pumping in a counterdirectional configuration. The Kerr effect permits to obtain high-beam quality amplification with no accompanying frequency conversions. As a result, our nonlinear taper amplifier may provide an important building block for multimode fiber lasers and amplifiers.

1. ## Introduction

Multimode optical fibers (MMFs) are currently the subject of renewed research interest, for their potential application in different optical technologies. For example, MMFs permit increasing the capacity of optical communications by means of spatial division multiplexing [1], up-scaling the energy from fiber lasers [2] and supercontinuum light sources [3,4], performing high-resolution biomedical imaging, and delivering powerful light beams for metal cutting and micromachining, to name a few.

Fiber lasers provide the most promising field of application of MMFs: the nonlinear transmission of a short span of graded-index (GRIN) MMF between singlemode fibers leads to an ultrafast saturable absorber mechanism with high damage threshold [5,6]. Moreover, nonlinear effects in MMF lasers may permit to combine both longitudinal and transverse mode-locking [7], and to achieve high power and simultaneously beam quality multimode laser sources based on spatial beam cleaning effects [8].

A special property of nonlinear MMFs is that they enable power-activated reshaping of the transverse spatial beam pattern at their output. This may occur by means of either stimulated scattering processes such as the Raman or Brillouin effects [8-11], or via the Kerr effect leading to spatial beam self-cleaning [12-15]. All of these effects lead to spatial beam cleaning, that is to the generation of a high quality output beam from the MMF, when pumped with low quality multimode pumps. In the absence of nonlinearities, the multitude of randomly excited fiber modes would lead to highly speckled or irregular intensity patterns at the MMF output.

On the other hand, it is well-known that the power scaling of high-power fiber lasers based on the use of rare-earth doped fibers as gain medium is limited by the presence of nonlinear effects such as Raman and Brillouin scattering [16-18], as well as by transverse mode instability (TMI) [19,20]. Particularly, TMI is a main limiting factor in large mode area (LMA) CW fiber lasers, leading to both sudden fluctuations and severe quality reduction of the beam at the amplifier output, for powers about a certain threshold value [17].

A possible solution to mitigate TMI is provided by the use of tapered ytterbium-doped fibers (T-YDFs), where the fiber core size varies along its axis, so that high-order modes (HOMs) are filtered out, and a good beam quality is obtained at their output. At the same time, the progressively increasing mode area permits to mitigate nonlinear effects, such as frequency conversions, which could otherwise distort and deplete the laser beam. Several researchers have reported the use of T-YDFs for single-mode fiber lasers operating in both pulsed and CW regimes [21-23], and a detailed intercomparison of the beam quality performance by using both uniform and tapered YD fiber amplifiers has been recently carried out [24]. In all of these studies, the T-YDFs are singlemode at their input end with a smaller diameter, so that quasi singlemode operation is maintained throughout their large diameter output end, with relatively small bending loss sensitivity, and high TMI thresholds when compared with uniform YDFs. However, for the development of multimode fiber lasers, which would permit a further up-scaling of the output power, multimode T-YDFs should be used.

In order to overcome the beam quality degradation that is inherent to active MMFs, nonlinearity may turn from a disadvantage into an opportunity via the beam self-cleaning effect. Guenard et al. demonstrated self-cleaning in a 5-m long double-clad uniform YD MMF with nearly step-index refractive index profile and uniform core doping profile [25]. Beam self-cleaning was also exploited to obtain mode-locking in a composite cavity laser configuration, where a passively Q-switched Nd:YAG microchip laser was combined with an extended cavity including the same YD MMF [26].

More recently, Niang et al. demonstrated Kerr beam self-cleaning in a 10-m long T-YD MMF, with parabolic core refractive index profile, using sub-nanosecond 1064 nm pulses propagating from the wider (120 µm) into the smaller (40 µm) diameter [27]. In this case, however, self-cleaning in the presence of amplifier gain was accompanied by ultrawideband frequency conversion and supercontinuum generation extending from 520 nm to 2600 nm. In fact, the decreasing core diameter in combination with amplification concur to accelerate the nonlinear effects, similar to the case of lossless singlemode [28] and multimode fiber tapers [29].

In order to use the T-YD MMF in a high power fiber laser, any frequency conversion should be avoided, so that the signal must necessarily propagate from the smaller to the larger diameter, similar to conventional T-YDF amplifiers [21-24]. In this configuration, which decreases the fiber nonlinearity as the pulse get amplified, it remains an open issue whether any beneficial Kerr-induced beam self-cleaning may still be achieved. In this work, we answer this question, by experimentally demonstrating that indeed beam self-cleaning is obtained in a ~5 m long T-YD MMFs, in spite of a longitudinally increasing core diameter, which leads to a decelerating nonlinearity. We used 500 ps pulses at 1064 nm, propagating in the normal dispersion regime, from the smallest to the largest taper diameter. Remarkably, we observed self-cleaning with its associated beam quality improvement not only in the active configuration, that is in the presence of gain (induced by the pump laser diode), but even in a lossy configuration (i.e., with no pump laser diode).

## 2. Experimental Details

The experimental setup is schematically illustrated in Fig. 1. We injected in the T-YD MMF a Nd:YAG microchip laser at 1064 nm with Gaussian spatial beam shape, generating 500 ps pulses at the repetition rate of 500 Hz, with up to 90 kW peak power. A polarizing beam-splitter (PBS) and two half-wave plates (HWP1,2) were used to adjust the input power and polarization state of the laser beam.

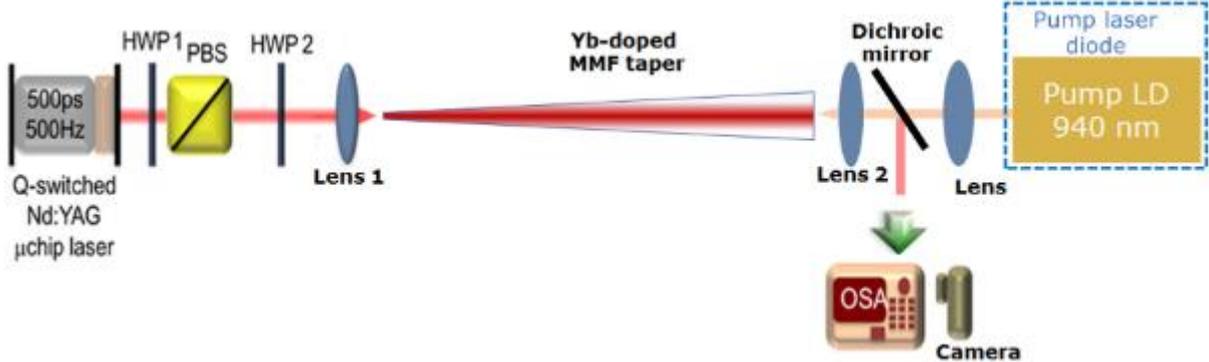

*Fig. 1. Schematic of the experimental setup for coupling microchip laser pulses at 1064 nm as well as a CW pump laser diode (LD) into the Yb-doped MMF taper.*

In experiments, we used a 5 m long Yb-doped MMF taper with 1.1 dB/m linear average attenuation at 1064 nm, corresponding to an overall attenuation of 5.5 dB, and parabolic core refractive index, with nearly uniform doping profile (for details, see [27]). The taper was wound on a fiber coil (not presented in Fig. 1). The smallest (largest) core diameter was close to 47 (96) µm (the corresponding cladding was square with 114 (275) µm side). Tapering the YD MMF diameter was obtained by controlling the winding speed of the fiber during drawing, leading to a nearly exponentially increasing core diameter along its length [27].

To pump the rare-earth Yb ions, we used a 10 W output power CW multimode laser diode (LD) (Keopsys, Lannion, France) operating at 940 nm, and coupled into a multimode fiber pigtail with 105 µm diameter and 0.22 numerical aperture. The taper was placed between two lenses. The first lens (Lens1) with a focal length of 31.4 mm was placed on a three-axis translation stage, for coupling the signal (with a beam diameter of 25 µm at full width of half maximum intensity (FWHMI)) into the smallest core side of the taper. Whereas the pump LD was coupled into the opposite, large core side of the taper, with a second lens (Lens2) of 35 mm focal length. The pump was focused into both the core and the cladding, so that pump light is also guided by the cladding of the double-clad YD MMF. To control the input coupling conditions of the signal beam, and to analyze the spatial output beam, the near-field beam profile from the large end face of the taper was imaged onto a CCD camera (Gentec Beamage-CCD12). We also used an optical spectrum analyzer (OSA) (Anritsu MS9710C: 600-1750 nm) in order to monitor the output spectrum.

## 3. Results

We experimentally studied the dependence of the output beam quality and the occurrence of beam self-cleaning in the T-YD MMF as a function of the peak power of the injected signal pulses. We carried out the study both in the passive case, that is in the absence of the pump, and in the active case, that is with the pump LD turned on.

### 3.1. Self-cleaning in lossy taper

The first experiment was performed in a passive configuration. The signal beam was focused into the smallest core of the T-YD MMF. The peak power of the coupled input signal ($P_{in}$) was initially set to 280 W: in this case a highly speckled output beam was observed, see top figure in the left panel of Fig. 2. As can be seen, when increasing the input coupled power above 20 kW, the output beam spatial pattern self-organises into a bright spot with a bell shape, surrounded by a low-power background, a typical manifestation of spatial beam self-cleaning [13]. The self-cleaned beam is stably maintained when the signal power grows up to 51 kW, which is the damage threshold of the taper input face. The right panels in Fig. 2 show that beam self-cleaning is not accompanied by any significant spectral broadening, nor frequency conversion processes. Although it is not shown in Fig. 2, we checked the absence of any stimulated Raman scattering induced Stokes peak around 1.1 µm. The occurrence of beam self-cleaning in a passive taper is a remarkable feature of multimode beam propagation, given that the beam experiences both significant linear loss and a decelerating nonlinearity, due to the widening of the core size along the taper. On the other hand, although the taper becomes increasingly multimodal as the beam propagates through it, the light beam gets progressively more coupled into the fiber core along the longitudinal direction, which facilitates the beam cleaning process.

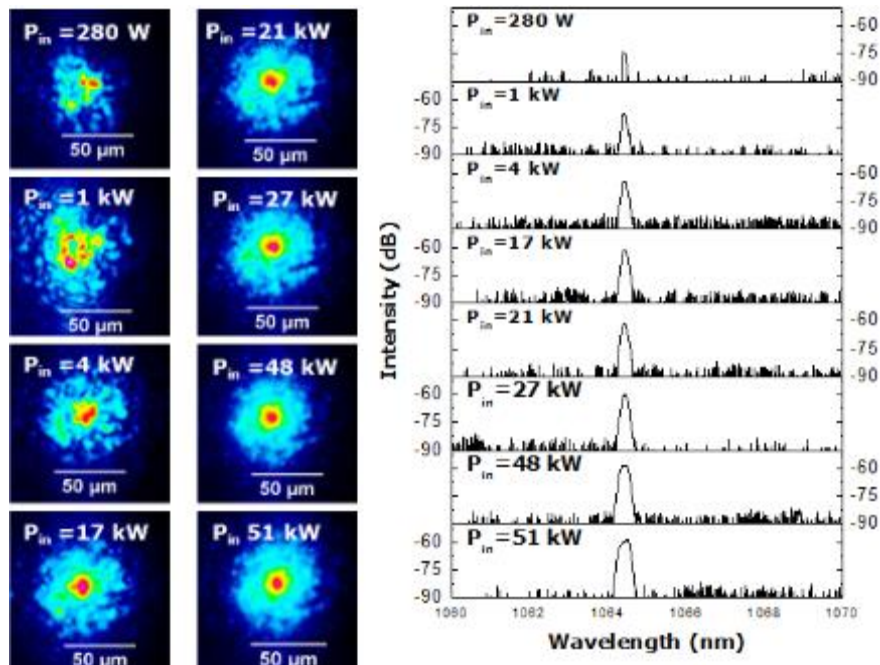

*Fig. 2. Left panels: Near-field spatial distributions; Right panels: spectra from the T-YD MMF output, versus input peak power ($P_{in}$) and without LD pump. The tapered fiber length was 5 m.*

As shown in Fig. 3, the increase of the output beam quality as a function of the input peak power was confirmed by the measurement of the $M^2$ beam quality parameter. As can be seen, the beam quality parameter drops from $M^2=12$ to below $M^2=4$ as the peak power grows above 20 kW, and then stabilises around $M^2=3$ when the power grows up to 51 kW. In addition, we confirmed a strong stability of the self-cleaned beam in the presence of bending of the taper, as it occurs in the lossless case [13].

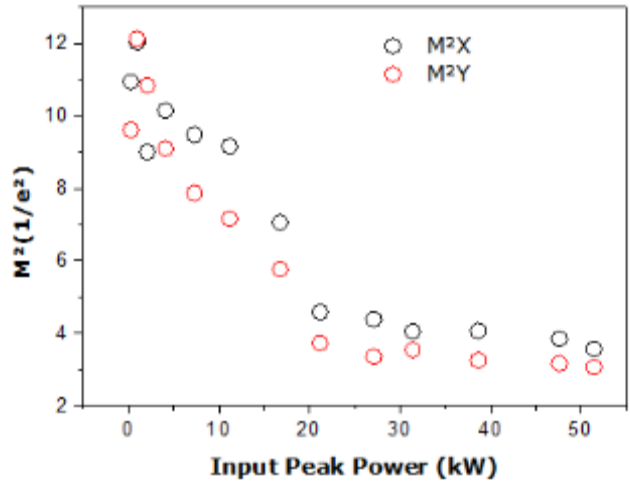

*Fig. 3. Measured beam quality $M^2$ parameter for orthogonal transverse dimensions versus input peak power, for the case of Fig. 2.*

### 3.2. Self-cleaning in active taper

Next, we switched on the CW LD pump source for inducing amplification in the T-YD MMF. The input beam size at the smallest taper face remained fixed at 25 µm, as in the passive configuration. First, we characterised the MMF amplifier by studying the dependence of the output amplified spontaneous emission (ASE) noise power on LD pump power. The left plot in Fig. 4 compares the dependence on LD pump power of the ASE noise power (measured in a spectral window of 10 nm around 1064 nm), as obtained from either the smallest or largest core diameter end face of the taper. As can be seen, the highest ASE noise power is obtained from the large core diameter end face.

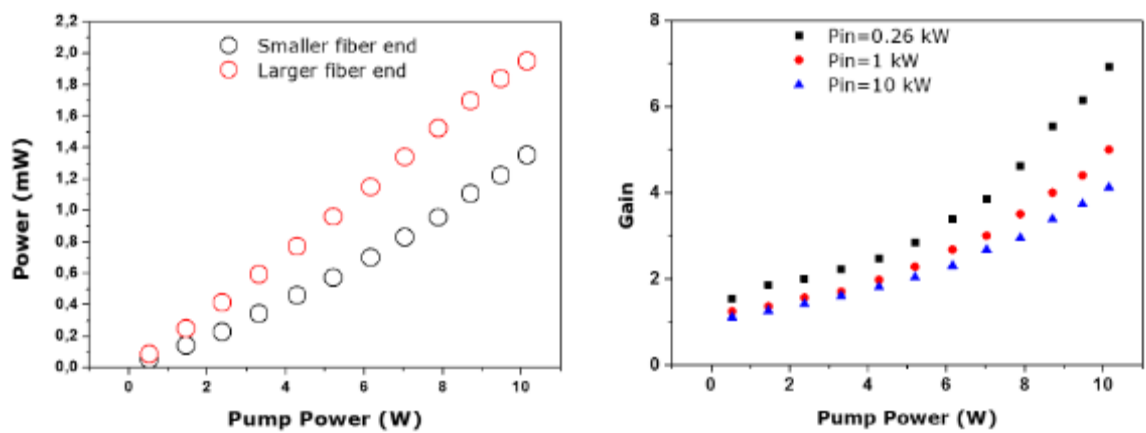

*Fig. 4. Left plot: amplification of ASE noise as a function of LD pump power; Right plot: net signal gain G vs. LD pump power, for different signal peak powers.*

On the other hand, the right plot of Fig. 4 illustrates the saturation of the gain G for the 1064 nm signal pulses, that we injected into the smallest core diameter end of the taper. Here the gain G is defined as the ratio between output peak power with pump, and output peak power without pump. Peak powers are measured at 1064 nm, by using a fast oscilloscope (6 GHz) coupled with a 5 GHz photodiode. Fig. 4 reveals that G≈7 for pump powers up to 10 W and signal powers around 260 W. Whereas amplification is strongly saturated, being limited to G≈ 4 as the signal peak power grows up to 10 kW. The gain G could be increased with a higher pump power in the counterpropagative configuration.

In spatial beam cleaning experiments, we kept the signal input peak power fixed at 10 kW, that is below the self-cleaning threshold in the passive configuration. Thus, the amplification boosts the signal and push it to enter the self-cleaning process.
The beam intensity patterns shown in Fig. 5 have been obtained by inserting a bandpass filter centered at 1064 nm with 10 nm bandwidth placed in front of the camera, to reduce the impact of ASE, and block residual back-reflected radiation from the pump. Fig. 5 (left panels) shows that the output field remains highly speckled, hence with low beam quality, until the gain reaches the value of 1.6. Conversely, for gain values above 1.6, the output beam remains stably self-cleaned.

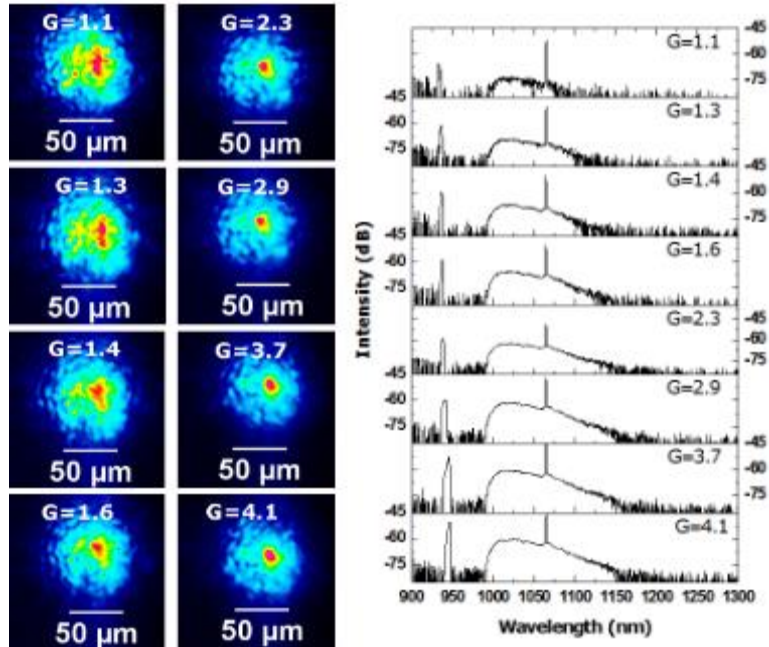

*Fig. 5. Left panels: Near-field spatial beam distribution from the largest core diameter face of the 5 m long taper, with 10 kW input power at 1064 nm, and different values of gain G; Right panels: corresponding spectra versus gain G.*

The output spectra corresponding to the manifestation of spatial beam self-cleaning are reported in the right panels of Fig. 5. As can be seen, also in the active case, Kerr-induced beam cleaning is not accompanied by any significant spectral broadening of the signal. This is in stark contrast with the case when the signal is injected at the large core side of the same taper: in that case, the combination of pulse amplification and accelerating nonlinearity due to the exponential decrease of the core diameter lead to octave-spanning supercontinuum generation, so that the signal at 1064 nm is substantially depleted [27]. Fig. 5 shows, in addition to the spectral line pertaining to the back-reflected 940 nm LD pump, a wideband ASE background surrounding the relatively narrow spectral line of the signal. The signal to ASE background noise power ratio is in a range of 15-20 dB, depending on the gain value.

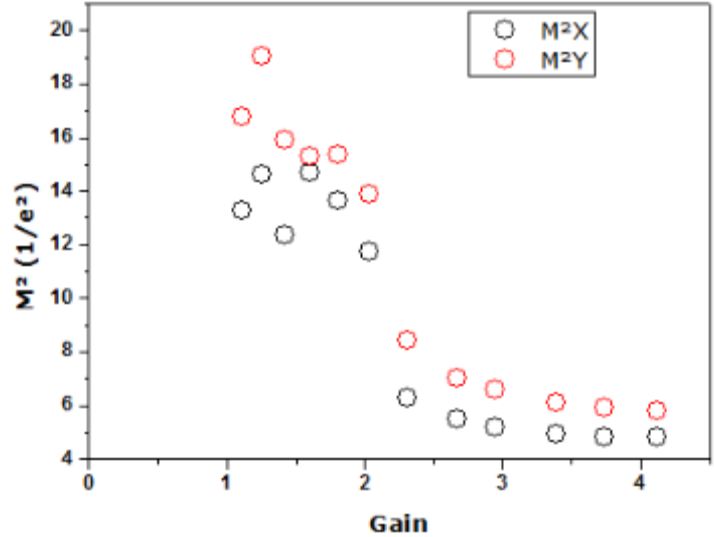

*Fig. 6. Measured beam quality $M^2$ parameter versus laser gain, for the case of Fig. 5.*

In the active case, the self-cleaning of the beam injected from the smaller side of the T-YD MMF was also characterised in terms of $M^2$ output beam quality parameter measurements, as shown in Fig. 6. Here, the beam quality parameter drops from very high values ($M^2=15$) for G=1.1, to about $M^2=5$ for the maximum gain value of 4.1. For the beam quality measurements, we used a 3 nm bandwidth filter at 1064 nm to further reduce the impact of ASE on the low power background. In fact, the $M^2$ parameter is strongly degraded by the presence of a wide noise-like background, as revealed in the output beam intensity profiles of Fig. 5. Similarly to the passive case, the self-cleaned beam remains highly insensitive to fiber bending induced perturbations. Note that, in our experiments, we focused our main attention on the demonstration of spatial self-cleaning. Many other fiber parameters, such as doping concentration, law of variation of the core size along the taper and its length, pump wavelength, etc., have not been optimized for maximizing gain.

## 4. Conclusions

To conclude, we experimentally demonstrated that tapered ytterbium-doped multimode fibers with a GRIN profile are suitable for Kerr-activated beam self-cleaning in both passive and active configurations, in spite of the decelerating nonlinearity due to an exponentially increasing core size. In both passive and active cases, obtaining spatial beam self-cleaning, without accompanying self-phase modulation induced spectral broadening or frequency conversion, constitutes a remarkable property of nonlinear multimode propagation in multimode tapers. Specifically, it results because of the occurrence of Kerr self-cleaning, in spite of a progressively decreasing nonlinearity. Self-cleaning turns out to be assisted by the widening of the core size along the propagation direction, which helps progressively trapping the beam into the fundamental fiber mode. Thanks to the presence of beam self-cleaning, multimode tapered amplifiers could provide an important building block for a future generation of high beam quality, and high power multimode fiber lasers and amplifiers.

**Acknowledgements**

We acknowledge support from the European Research Council (ERC) under the European Union's Horizon 2020 research and innovation program (grant No. 740355), and by the Russian Ministry of Science and Education, Grant No 14.Y26.31.0017. XLIM has received funding from Agence Nationale de la Recherche (ANR) (TRAFIC

**References**

[1] D. J. Richardson, J. M. Fini, and L. Nelson, "Space-division multiplexing in optical fibres," Nat. Photon. 7, 354–362 (2013).
[2] W. Fu, L. G. Wright, P. Sidorenko, S. Backus, and F. W. Wise, "Several new directions for ultrafast fiber lasers," Opt. Express 26, 9432–9463 (2018). [Online]. Available: http://www.opticsexpress.org/abstract.cfm?URI=oe-26-8-9432
[3] G. L. Galmiche, Z. S. Eznaveh, M. A. Eftekhar, J. A. Lopez, L. G. Wright, F. Wise, D. Christodoulides, and R. A. Correa, "Visible supercontinuum generation in a graded index multimode fiber pumped at 1064 nm," Opt. Lett. 41, no. 11, 2553–2556, (2016).
[4] K. Krupa, C. Louot, V. Couderc, M. Fabert, R. Guenard, B. M. Shalaby, A. Tonello, D. Pagnoux, P. Leproux, A. Bendahmane, R. Dupiol, G. Millot, and S. Wabnitz, "Spatiotemporal characterization of supercontinuum extending from the visible to the mid-infrared in a multimode graded-index optical fiber," Opt. Lett. 41, 5785–5788 (2016).
[5] E. Nazemosadat and A. Mafi, "Nonlinear multimodal interference and saturable absorption using a short gradedindex multimode optical fiber," J. Opt. Soc. Am. B 30, 1357–1367 (2013).
[6] S. Fu, Q. Sheng, X. Zhu, W. Shi, J. Yao, G. Shi, R. A. Norwood, and N. Peyghambarian, "Passive Q-switching of an all-fiber laser induced by the Kerr effect of multimode interference," Opt. Express 23, 17255–17262, (2015). [Online]. Available: http://www.opticsexpress.org/abstract.cfm?URI=oe-23-13-17255
[7] L. G. Wright, D. N. Christodoulides, and F. W. Wise, "Spatiotemporal mode-locking in multimode fiber lasers," Science 358, 94–97 (2017).
[8] E. A. Zlobina, S. I. Kablukov, A. A. Wolf, I. N. Nemov, A. V. Dostovalov, V. A. Tyrtyshnyy, D. V. Myasnikov, and S. A. Babin, "Generating high-quality beam in a multimode LD-pumped all-fiber Raman laser," Opt. Express 25, 12 581–12 587 (2017). [Online]. Available: http://www.opticsexpress.org/abstract.cfm?URI=oe-25-11-12581
[9] K. S. Chiang, "Stimulated Raman scattering in a multimode optical fiber: evolution of modes in Stokes waves," Opt. Lett. 17, 352–354, (1992). [Online]. Available: http://ol.osa.org/abstract.cfm?URI=ol- 17-5-352
[10] H. Bruesselbach, "Beam cleanup using stimulated Brillouin scattering in multimode fibers," in Conference on Lasers and Electro-Optics. Optical Society of America (1993), p. CThJ2. [Online]. Available: http://www.osapublishing.org/abstract.cfm?URI=CLEO-1993-CThJ2
[11] L. Lombard, A. Brignon, J.-P. Huignard, E. Lallier, and P. Georges, "Beam cleanup in a self-aligned gradient-index Brillouin cavity for high-power multimode fiber amplifiers," Opt. Lett. 31, 158–160 (2006). [Online]. Available: http://ol.osa.org/abstract.cfm?URI=ol-31-2-158
[12] K. Krupa, A. Tonello, A. Barthélémy, V. Couderc, B. M. Shalaby, A. Bendahmane, G. Millot, and S. Wabnitz, "Observation of geometric parametric instability induced by the periodic spatial self-imaging of multimode waves," Phys. Rev. Lett. 116,183901 (2016).
[13] K. Krupa, A. Tonello, B. M. Shalaby, M. Fabert, A. Barthélémy, G. Millot, S. Wabnitz, and V. Couderc, "Spatial beam self-cleaning in multimode fibres," Nat. Photon. 11, 234–241 (2017).
[14] Z. Liu, L. G. Wright, D. N. Christodoulides, and F. W. Wise, "Kerr self-cleaning of femtosecond-pulsed beams in graded-index multimode fiber," Opt. Lett. 41, 3675–3678 (2016).
[15] L. G. Wright, Z. Liu, D. A. Nolan, M.-J. Li, D. N. Christodoulides, and F. W. Wise, "Self-organized instability in graded-index multimode fibres," Nat. Photon. 10, 771–776, (2016).
[16] J. Nilsson and D. N. Payne, "High-power fiber lasers," Science 332, 921–922 (2011). [Online]. Available: https://science.sciencemag.org/content/332/6032/921
[17] C. Jauregui, J. Limpert, and A. Tünnermann, "High-power fibre lasers," Nat. Photon. 7, 861–867 (2013).
[18] M. N. Zervas and C. A. Codemard, "High power fiber lasers: A review," IEEE Journal of Selected Topics in Quantum Electronics, 20, 219–241, (2014).
[19] T. Eidam, C. Wirth, C. Jauregui, F. Stutzki, F. Jansen, H.-J. Otto, O. Schmidt, T. Schreiber, J. Limpert, and A. Tünnermann, "Experimental observations of the threshold-like onset of mode instabilities in high power fiber amplifiers," Opt. Express 19, 13218–13224 (Jul 2011). [Online]. Available: http://www.opticsexpress.org/abstract.cfm?URI=oe-19-14-13218
[20] C. Jauregui, T. Eidam, J. Limpert, and A. Tünnermann, "Impact of modal interference on the beam quality of high-power fiber amplifiers," Opt. Express 19, 3258–3271 (2011). [Online]. Available: http://www.opticsexpress.org/abstract.cfm?URI=oe-19-4-3258
[21] L. Li, Q. Lou, J. Zhou, J. Dong, Y. Wei, S. Du, and B. He, "High power single transverse mode operation of a tapered large-mode-area fiber laser," Opt. Commun 281, 655–657 (2008). [Online]. Available: http://www.sciencedirect.com/science/article/pii/S0030401807009765
[22] J. Kerttula, V. Filippov, Y. Chamorovskii, V. Ustimchik, K. Golant, and O. G. Okhotnikov, "Tapered fiber amplifier with high gain and output power," Laser Physics 22, 1734–1738 (2012). [Online]. Available: https://doi.org/10.1134/S1054660X12110059
[23] K. Bobkov, A. Andrianov, M. Koptev, S. Muravyev, A. Levchenko, V. Velmiskin, S. Aleshkina, S. Semjonov, D. Lipatov, A. Guryanov, A. Kim, and M. Likhachev, "Sub-mw peak power diffraction-limited chirped-pulse monolithic yb-doped tapered fiber amplifier," Opt. Express 25, 26 958–26 972 (2017). [Online]. Available: http://www.opticsexpress.org/abstract.cfm?URI=oe-25-22-26958
[24] Y. Ye, X. Xi, C. Shi, B. Yang, X. Wang, H. Zhang, P. Zhou, and X. Xu, "Comparative study on transverse mode instability of fiber amplifiers based on long tapered fiber and conventional uniform fiber," Laser Physics Letters 16, 085109 (2019).
[25] R. Guenard, K. Krupa, R. Dupiol, M. Fabert, A. Bendahmane, V. Kermene, A. Desfarges-Berthelemot, J. L. Auguste, A. Tonello, A. Barthélémy, G. Millot, S. Wabnitz, and V. Couderc, "Kerr self-cleaning of pulsed beam in an ytterbium doped multimode fiber," Opt. Express, 25, 4783–4792 (2017).
[26] R. Guenard, K. Krupa, R. Dupiol, M. Fabert, A. Bendahmane, V. Kermene, A. Desfarges-Berthelemot, J. L. Auguste, A. Tonello, A. Barthélémy, G. Millot, S. Wabnitz, and V. Couderc, "Nonlinear beam self-cleaning in a coupled cavity composite laser based on multimode fiber," Opt. Express 25, 22219-22227 (2017).
[27] A. Niang, T. Mansuryan, K. Krupa, A. Tonello, M. Fabert, P. Leproux, D. Modotto, O. N. Egorova, A. E. Levchenko, D. S. Lipatov, S. L. Semjonov, G. Millot, V. Couderc, and S. Wabnitz, "Spatial beam self-cleaning and supercontinuum generation with yb-doped multimode graded-index fiber taper based on accelerating self-imaging and dissipative landscape," Opt. Express 27, 24018–24028 (2019). [Online]. Available: http://www.opticsexpress.org/abstract.cfm?URI=oe-27-17-24018
[28] T. A. Birks, W. J. Wadsworth, and P. S. J. Russell, "Supercontinuum generation in tapered fibers," Opt. Lett. 25, 1415–1417 (2000). [Online]. Available: http://ol.osa.org/abstract.cfm?URI=ol-25-19-1415
[29] M. A. Eftekhar, Z. Sanjabi-Eznaveh, H. E. Lopez-Aviles, S. Benis, J. E. Antonio-Lopez, M. Kolesik, F. Wise, R. Amezcua-Correa, and D. N. Christodoulides, "Accelerated nonlinear interactions in gradedindex multimode fibers," Nat. Commun.10, 1638 (2019). [Online]. Available: https://doi.org/10.1038/s41467-019-09687-9